\PassOptionsToPackage{table}{xcolor}
\documentclass[conference]{IEEEtran}

\usepackage{cite}
\usepackage{amsmath,amssymb,amsfonts}
\usepackage{graphicx}
\usepackage{textcomp}
\usepackage{xcolor}
\usepackage{tabularx}
\usepackage{array}
\usepackage{calc}
\usepackage{multirow}
\usepackage{booktabs}
\usepackage{threeparttable}
\usepackage{tikz}
\usetikzlibrary{positioning,calc,arrows.meta,shadows.blur,decorations.pathreplacing}
\usepackage{tabularray}
\UseTblrLibrary{booktabs}
\usepackage{enumitem}
\usepackage{hyperref}

\definecolor{qbffBlue}{HTML}{3568DD}
\definecolor{qbffGreen}{HTML}{57944C}
\definecolor{qbffPurple}{HTML}{8450D8}
\definecolor{qbffOrange}{HTML}{DC7F22}
\definecolor{qbffMagenta}{HTML}{BC356B}

\AtBeginDocument{%
}

\begin{document}

\title{An Initial Study of Physiological Markers for Deepfake Speech Detection}

% All authors share one affiliation. The long IEEE conference author format
% keeps the required author order while avoiding repeated affiliation blocks.
\author{
\IEEEauthorblockN{Vojtěch Staněk, Michal Frič, Kamil Malinka, and Anton Firc}
\IEEEauthorblockA{Security@FIT, Brno University of Technology\\
Brno, Czech Republic\\
istanek@fit.vut.cz, xfricm02@vut.cz, malinka@fit.vut.cz, ifirc@fit.vut.cz}
}

\maketitle

\begin{abstract}
Conventional deepfake speech detectors primarily target synthesis artifacts and often degrade under unseen attacks or recording conditions. Human speech, however, contains physiological markers associated with respiration, phonation, and temporal coordination arising from the speech-production system that synthesizers may not reproduce faithfully. This motivates the hypothesis that deepfake speech can be detected by identifying inconsistencies in these patterns, which we test using 36 handcrafted physiological markers organized into respiratory, vocal, and temporal categories. We evaluate the marker vectors with logistic regression and SVM classifiers on pooled English recordings. The best in-domain system achieves a $21.40\%$ equal error rate (EER), while logistic regression achieves $32.77\%$ on the out-of-domain dataset pair. Since the markers describe specific speech-production properties, the logistic-regression coefficients indicate whether higher marker values push the classifier toward bona fide or deepfake speech. These results provide initial evidence that physiological markers offer a useful, interpretable signal for deepfake speech detection.
\end{abstract}

\begin{IEEEkeywords}
deepfake speech, physiological markers, explainability, speech production, audio forensics
\end{IEEEkeywords}

%%%%%%%%% INTRODUCTION
\section{Introduction}
\label{sec:intro}

% \vojta{rough draft hotov. Je potreba to jeste zkratit (cca 1.5-2 stranky) a prorezat reference, at se vlezeme do 8+2 stran}

The rapid advancement of text-to-speech (TTS) and voice conversion (VC) technologies has enabled effortless generation of convincing deepfake speech~\cite{wang24_asvspoof}.
These deepfakes pose a serious threat to automatic speaker verification (ASV) systems, enable fraud, and undermine trust in audio evidence~\cite{Firc2, MalinkaVoiceAssistants}.
To counter this threat, current deepfake speech detection systems typically rely on data-driven classifiers that learn to identify synthesis artifacts in the acoustic signal~\cite{Firc3}.
While highly effective, these systems often rely on patterns specific to the synthesizers, codecs, or recording conditions being examined, which can lead to poor out-of-domain generalization to unseen attacks or new conditions~\cite{muller2022inthewild}.

In contrast, human speech carries inherent physiological markers associated with respiration, phonation, and temporal coordination, originating in the human speech-production system~\cite{elgiriyewithana2024breathing}.
These markers are difficult to reproduce faithfully, even in modern
state-of-the-art synthesizers~\cite{xtts}.
This motivates the core hypothesis of this work: \textit{deepfake speech can be detected by identifying inconsistencies in the physiological patterns of natural speech production}.
We therefore propose a biologically grounded detector based on 36 handcrafted physiological markers capturing respiratory, vocal, and temporal properties.

Unlike conventional detectors based on high-dimensional learned representations, the proposed approach operates in a semantically meaningful feature space.
Combined with a simple linear classifier, it can show how individual speech-production markers contribute to its detection score.
On a pooled corpus of English recordings %longer than 14\,s
drawn from several modern datasets, the standalone detector achieves an Equal Error Rate (EER) of $21.40\%$ in-domain and $32.77\%$ under out-of-domain evaluation.
These results provide initial evidence that physiologically grounded markers contain a usable and interpretable signal for deepfake speech detection. In this paper, we make the following contributions:
\begin{enumerate}[topsep=-4pt, itemsep=3pt]
    % \item We propose and implement an interpretable deepfake speech detection approach based on 36 handcrafted physiological markers capturing respiratory, vocal, and temporal properties of speech production.

    \item We propose and implement an intrinsically interpretable, low-dimensional deepfake speech detector based on 36 handcrafted markers spanning respiratory, vocal, and temporal properties of speech production.
    
    \item We evaluate the proposed detector under both in-domain and out-of-domain conditions and demonstrate that it provides a measurable standalone detection signal.
    
    \item We show that the coefficients of a linear classifier provide intrinsic marker-level explanations, revealing which characteristics support bona fide or deepfake decisions.
\end{enumerate}

%-------------------------------------------------------------------------
\section{Related Work}
\label{sec:related}

% \subsection{Deepfake Speech Detection}

\textbf{Deepfake speech detection} has been addressed using both handcrafted acoustic features and neural architectures~\cite{Firc3}. 
Early approaches relied on spectral and cepstral representations, such as MFCCs~\cite{Liu2021} or CQCCs~\cite{todisco2019asvspoof}, to capture synthesis artifacts.
More recent systems increasingly employ self-supervised learning (SSL) feature extractors~\cite{Babu2021, tak2022} combined with dedicated processing and detection modules~\cite{stanek2026evolutionary} such as AASIST~\cite{Jung2022aasist}, MHFA~\cite{rohdin24_asvspoof}, or simple mean pooling~\cite{stanek2026evolutionary}. 
These approaches achieve strong performance on benchmark datasets~\cite{todisco2019asvspoof, wang24_asvspoof}. %primarily target artifacts introduced by specific synthesis pipelines.
A persistent challenge in this area is generalization to unseen attacks and recording conditions; prior work has shown that detectors trained on known synthesis methods may degrade substantially when evaluated on out-of-domain data, suggesting that many systems rely on cues tied to specific generators, codecs, or conditions~\cite{muller2022inthewild}.
%A further limitation of some current detectors is that they are designed to operate on fixed-length input segments, typically around 4\,s~\cite{chan_4s, Zhang24SLS}. 
% While effective for artifact-based detection on standard benchmarks, such a design limits the model's temporal context and may hinder the capture of longer-range physiological patterns, such as breathing behavior~\cite{longform_audio_study}.

\newcommand{\mcell}[1]{\parbox[c]{\widthof{Harmonics-to-noise ratio (HNR)}}{\raggedright #1}}

\begin{table*}[t]
\centering
\begin{threeparttable}
\caption{Physiologically motivated categorization of the handcrafted markers in this work.}
\label{tab:markers}
\small
\begin{tabularx}{\textwidth}{@{}
>{\raggedright\arraybackslash}m{\widthof{\textbf{Respiratory}}}
>{\raggedright\arraybackslash}m{\widthof{Harmonics-to-noise ratio (HNR)}}
>{\raggedright\arraybackslash}X
@{}}
\toprule
\textbf{Category} & \textbf{Marker} & \textbf{Measures} \\
\midrule
\multirow{5}{*}[-2.5ex]{\shortstack[l]{\textbf{Respiratory}\\(14 markers)}}
& \mcell{Breath frequency} & frequency \\
& \mcell{Breath duration} & mean, std, min, max \\
& \mcell{Breath acoustics} & spectral centroid, spectral bandwidth, spectral peak frequency \\
& \mcell{Pre-speech inhalation} & pre-speech inhalation ratio \\
& \mcell{Speech-breath coordination} & coordination score, speech-to-breath mean/std, breath-to-speech mean/std \\
\midrule
\multirow{4}{*}{\shortstack[l]{\textbf{Vocal}\\(16 markers)}}
& \mcell{Fundamental frequency ($F_0$)} & mean, std, min, max \\
& \mcell{Jitter} & absolute, relative, RAP, PPQ5\tnote{$\dagger$} \\
& \mcell{Shimmer} & absolute, dB, relative, APQ3, APQ5, APQ11\tnote{$\ddagger$} \\
& \mcell{Harmonics-to-noise ratio (HNR)} & mean, std \\
\midrule
\multirow{3}{*}[-1.5ex]{\shortstack[l]{\textbf{Temporal}\\(6 markers)}}
& \mcell{Speech tempo} & speech rate, articulation rate \\
& \mcell{Pause structure} & mean pause duration, pause frequency, total pause duration \\
& \mcell{Filled pauses} & filler word frequency \\
\bottomrule
\end{tabularx}
\begin{tablenotes}[flushleft]
\footnotesize
\item[$\dagger$] RAP and PPQ5 are jitter measures based on relative average perturbation and five-point period perturbation quotient, respectively.
\item[$\ddagger$] APQ3, APQ5, and APQ11 are shimmer measures based on amplitude perturbation quotients computed over 3, 5, and 11 periods.
\end{tablenotes}
\end{threeparttable}
% \vspace{-1.3em}
\end{table*}

A common thread across these methods is their focus on acoustic artifacts introduced during synthesis. In contrast, our work investigates biologically grounded characteristics motivated by speech production rather than explicit modeling of synthesis artifacts.

% \subsection{Physiological Cues in Speech Analysis}

% The speech production process involves complex coordination of respiratory, phonatory, and articulatory systems, 
% each leaving characteristic traces in the acoustic signal.
% Breath detection has received attention with neural models such as Respiro~\cite{}, which demonstrated that breathing patterns carry speaker-specific information.
% Vocal quality measures - jitter, shimmer, and harmonics-to-noise ratio - are well-established in clinical speech analysis~\cite{} for assessing laryngeal pathologies and voice disorders.
% Temporal speech characteristics, including pause patterns and speech rate variability, have been linked to cognitive planning processes~\cite{} and used in speaker profiling.

In parallel, prior work has examined \textbf{physiological aspects of human speech production}.
Breathing is tightly coupled with speech planning and utterance structure, and both breathing patterns and breath sounds have been shown to carry speaker-related information~\cite{tran2023breath}. Breath-based deepfake detection has recently been explored by Layton et al.~\cite{layton_breath}, who derive global statistics from detected breath events, and by BreathNet~\cite{breathnet}, which integrates breath cues into an SSL-based detector to improve cross-benchmark generalization.
In contrast, our objective is a systematic and intrinsically interpretable analysis of a low-dimensional marker space spanning various physiological properties.
At the same time, vocal quality measures such as fundamental frequency, jitter, shimmer, and harmonics-to-noise ratio are well-established acoustic markers of phonatory stability and are widely used in clinical voice analysis~\cite{fantini2024voice}.
Temporal measures, including speech rate, articulation rate, and pause patterns, are also commonly used to describe speech fluency and production dynamics~\cite{angelopoulou2024fluency}. Despite their relevance to human speech production, these physiological aspects remain underexplored in deepfake speech detection. Motivated by this gap, we investigate whether handcrafted physiological markers of speech production can support deepfake speech detection and complement modern SSL-based approaches.

%-------------------------------------------------------------------------
\section{Proposed Method}
\label{sec:method}

We propose a biologically grounded deepfake speech detector based on handcrafted physiological markers of human speech production, motivated by the observation that bona fide speech inherently reflects coordinated physiological processes, whereas synthetic speech may fail to faithfully reproduce these patterns.
Rather than targeting conventional synthesis artifacts, our method aims to capture deviations from the physiological patterns of natural speech production.
Given an input speech recording, we extract a set of handcrafted physiological markers that describe various aspects of speech production.
These markers are then used as input to a linear classifier that predicts whether the recording is bona fide or deepfake.
Because several of the considered markers require sufficiently long temporal context, we restrict the analysis to English speech recordings longer than 14 seconds.
This design choice enables more reliable estimation of respiratory and temporal characteristics, which is critical for the proposed detector.
% \subsection{Physiological Markers}
\label{sec:taxonomy}
% Note: Seriozni taxonomie by byla banger rozsireni na zurnal potom...
To obtain a structured, physiologically meaningful representation, we organize the 36 handcrafted features into three categories that reflect different aspects of human speech production: \emph{respiratory}, \emph{vocal}, and \emph{temporal}, as visible in \autoref{tab:markers}.
This categorization is grounded in prior speech analysis literature, where breathing behavior~\cite{tran2023breath}, phonation patterns~\cite{fantini2024voice}, and speech fluency~\cite{angelopoulou2024fluency} have all been studied as informative channels of speech production and analysis. % This grouped representation is also useful for subsequent analysis, as it enables us to study the relative contribution of respiratory, vocal, and temporal information to deepfake speech detection.

%%%%%%%%%%%%%%%%%%%%%%%%%%%%

\textit{Respiratory markers} characterize breathing behavior and the coordination between breathing and speech.
This category includes measures related to breath frequency, breath duration, breath acoustics, pre-speech inhalation, and timing relationships between speech and breath.
Breathing is an integral part of natural speech. Speakers take breaths at linguistically appropriate locations, yet most TTS systems do not model it explicitly, which can result in synthetic speech with missing or artificially placed breath sounds~\cite{Li2020Breathing, szekely2020breathing, mostaani22_interspeech}.
% Such features aim to capture whether the recording exhibits breathing patterns consistent with natural human speech production.

\textit{Vocal markers} describe phonation characteristics, stability, and voice behavior. These features reflect biomechanical vocal properties of the human larynx that arise from the mechanics of voiced speech production.
Included are statistics of the fundamental frequency $F_0$, perturbation measures such as jitter and shimmer, along with voice harmonics~\cite{fantini2024voice}.

\textit{Temporal markers} capture the macro-level speech structure in time and describe features connected with speaking rhythm, speech fluency, and articulation timing patterns.
This category includes measures such as speech rate, articulation rate, and pause statistics (duration, frequency, placement). % as well as syllable-level measures and other timing-related cues.

\begin{figure*}[t]
\centering
\resizebox{0.8\textwidth}{!}{%
\begin{tikzpicture}[
    font=\small,
    >=Latex,
    col/.style={
        rounded corners=3mm,
        draw=#1!70!black,
        fill=#1!5,
        very thick,
        minimum height=5.25cm,
        inner sep=2mm
    },
    header/.style={
        rounded corners=2mm,
        draw=#1!70!black,
        fill=#1,
        text=white,
        font=\bfseries,
        align=center,
        minimum height=8mm
    },
    subbox/.style={
        rounded corners=2mm,
        draw=#1!70!black,
        fill=#1!10,
        thick,
        align=center,
        inner xsep=2mm,
        inner ysep=1.5mm,
        blur shadow={shadow blur steps=4}
    },
    markerbox/.style={
        subbox=#1,
        text width=2.70cm,
        minimum height=9mm
    },
    arrow/.style={
        -{Latex[length=3mm,width=2mm]},
        very thick,
        draw=black!70
    },
    innerarrow/.style={
        -{Latex[length=2.2mm,width=1.5mm]},
        thick,
        draw=black!60
    }
]

% ------------------------------------------------------------------
% Main columns
% ------------------------------------------------------------------
\node[col=qbffBlue, minimum width=4.5cm] (c1) {};

\node[
    col=qbffPurple,
    minimum width=5.5cm,
    right=5mm of c1
] (c2) {};

\node[
    col=qbffMagenta,
    minimum width=5cm,
    right=5mm of c2
] (c3) {};

% ------------------------------------------------------------------
% Headers
% ------------------------------------------------------------------
\node[
    header=qbffBlue,
    minimum width=4.5cm,
    anchor=north
] at (c1.north) {
    Input and preprocessing
};

\node[
    header=qbffPurple,
    minimum width=5.5cm,
    anchor=north
] at (c2.north) {
    Physiological markers
};

\node[
    header=qbffMagenta,
    minimum width=5cm,
    anchor=north
] at (c3.north) {
    Classification
};

% ------------------------------------------------------------------
% Input and preprocessing
% ------------------------------------------------------------------
\node[
    align=center,
    anchor=north,
    text width=3.1cm
] at ($(c1.north)+(0,-10mm)$) {
    \textbf{Waveform}\\[-0.3mm]
    duration $>14$\,s
};

% Waveform
\begin{scope}[shift={($(c1.north)+(0,-25mm)$)}]
\foreach \x/\h in {
    -1.15/0.40,
    -0.95/0.85,
    -0.75/1.10,
    -0.55/0.55,
    -0.35/0.80,
    -0.15/0.35,
     0.05/0.65,
     0.25/1.00,
     0.45/0.50,
     0.65/0.90,
     0.85/0.45,
     1.05/0.70
}{
    \draw[
        qbffBlue,
        line width=1.5pt,
        line cap=round
    ] (\x,-\h/2) -- (\x,\h/2);
}
\end{scope}

\node[
    subbox=qbffBlue,
    text width=2.9cm,
    anchor=south
] at ($(c1.south)+(0,3.5mm)$) {
    \textbf{Preprocessing}\\[0.4mm]
    16\,kHz mono\\
    Peak normalization
    % RMS silence mask
};

% ------------------------------------------------------------------
% Physiological marker extraction
% ------------------------------------------------------------------
\node[
    markerbox=qbffOrange,
    anchor=north west
] (resp) at ($(c2.north west)+(3mm,-11mm)$) {
    \textbf{Respiratory}\\[-0.2mm]
    \footnotesize 14 markers
};

\node[
    markerbox=qbffPurple,
    anchor=north
] (vocal) at ($(resp.south)+(0,-5mm)$) {
    \textbf{Vocal}\\[-0.2mm]
    \footnotesize 16 markers
};

\node[
    markerbox=qbffGreen,
    anchor=north
] (temp) at ($(vocal.south)+(0,-5mm)$) {
    \textbf{Temporal}\\[-0.2mm]
    \footnotesize 6 markers
};

% Input fork
% \coordinate (fork) at ($(c2.west)+(4mm,0)$);

% \draw[
%     thick,
%     black!55
% ] ($(fork|-resp.west)$) -- ($(fork|-temp.west)$);

% \draw[innerarrow]
%     ($(fork|-resp.west)$) -- (resp.west);

% \draw[innerarrow]
%     ($(fork|-vocal.west)$) -- (vocal.west);

% \draw[innerarrow]
%     ($(fork|-temp.west)$) -- (temp.west);

% Group the three categories
\coordinate (braceTop)
    at ($(resp.north east)+(2mm,0mm)$);

\coordinate (braceBot)
    at ($(temp.south east)+(2mm,0mm)$);

\draw[
    decorate,
    decoration={
        brace,
        amplitude=4.5pt,
        % mirror
    },
    thick,
    draw=black!70
] (braceTop) -- (braceBot);

\coordinate (braceMid)
    at ($(braceTop)!0.5!(braceBot)$);

% Combined marker representation
\node[
    subbox=qbffMagenta,
    text width=0.8cm,
    minimum height=2.0cm,
    anchor=east
] (repr) at ($(c2.east)+(-3mm,-3.5mm)$) {
    \begin{tikzpicture}[rotate=-90]
        \filldraw[
            draw=qbffOrange!75!black,
            fill=qbffOrange!25,
            rounded corners=0.35mm
        ] (0,0) rectangle (0.98,0.7);

        \node[
            overlay,
            anchor=center,
            inner sep=0pt,
            font=\scriptsize\bfseries
        ] at (0.49,0.35) {14};

        \filldraw[
            draw=qbffPurple!75!black,
            fill=qbffPurple!25,
            rounded corners=0.35mm
        ] (1.1,0) rectangle (2.22,0.7);

        \node[
            overlay,
            anchor=center,
            inner sep=0pt,
            font=\scriptsize\bfseries
        ] at (1.66,0.35) {16};

        \filldraw[
            draw=qbffGreen!75!black,
            fill=qbffGreen!25,
            rounded corners=0.35mm
        ] (2.32,0) rectangle (2.74,0.7);

        \node[
            overlay,
            anchor=center,
            inner sep=0pt,
            font=\scriptsize\bfseries
        ] at (2.53,0.35) {6};
    \end{tikzpicture}\\[1mm]
    36-D\\vector
};

% \draw[innerarrow]
%     ($(braceMid)+(1mm,0)$) -- (repr.west);

% ------------------------------------------------------------------
% Classification and explanation
% ------------------------------------------------------------------
\node[
    subbox=qbffBlue,
    text width=4.1cm,
    anchor=north
] (prep) at ($(c3.north)+(0,-11mm)$) {
    \textbf{Feature processing}\\[0.6mm]
    {\footnotesize Impute + standardize}
};

\node[
    subbox=qbffMagenta,
    text width=4.1cm,
    anchor=north
] (clf) at ($(prep.south)+(0,-4mm)$) {
    \textbf{Linear classifier}\\[0.6mm]
    \textit{deepfake score}
};

\node[
    subbox=qbffOrange,
    text width=4.1cm,
    anchor=north
] (exp) at ($(clf.south)+(0,-4mm)$) {
    \textbf{Marker-level interpretability}\\[0.6mm]
    {\footnotesize Coefficients reveal supporting cues}
};

\draw[innerarrow]
    (prep.south) -- (clf.north);

\draw[innerarrow] 
    (clf.south) -- (exp.north);

% ------------------------------------------------------------------
% Main pipeline arrows
% ------------------------------------------------------------------
\draw[arrow]
    (c1.east) -- (c2.west);

\draw[arrow]
    (c2.east) -- (c3.west);

\end{tikzpicture}%
}
\caption{
Overview of the proposed physiological-marker detector (best viewed in color).
The input recording is represented using respiratory,
vocal, and temporal markers.
The resulting 36-dimensional vector is processed by a linear classifier,
whose coefficients provide marker-level interpretability.
}
\label{fig:pipeline}
\end{figure*}
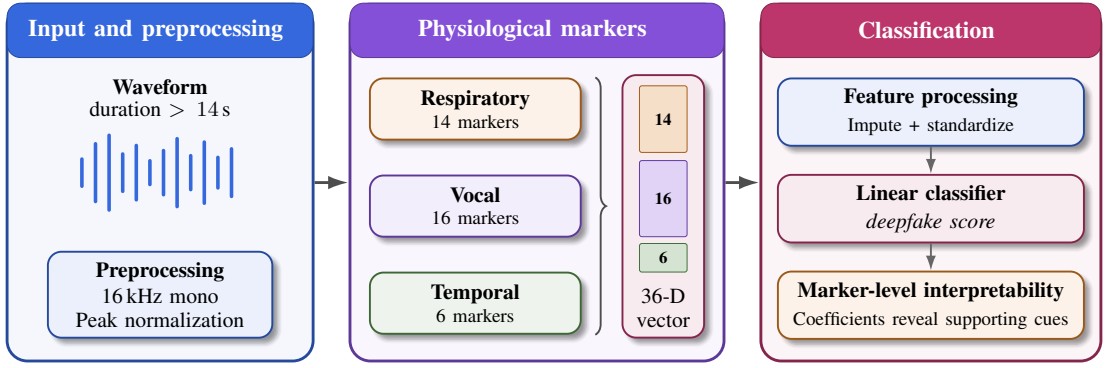

Natural speech exhibits characteristic pauses and disfluencies caused by cognitive speech planning, which synthesizers typically do not replicate~\cite{angelopoulou2024fluency, deJong2021fluency}.

% \subsection{Feature Extraction Pipeline}
% \label{sec:extraction}

% Required TikZ libraries:
% \usetikzlibrary{
%     positioning,
%     calc,
%     arrows.meta,
%     shadows.blur,
%     decorations.pathreplacing
% }

\noindent \textbf{Feature Extraction Pipeline} \quad
An overview of the proposed pipeline is shown in \autoref{fig:pipeline}. Given an input recording, the system converts the audio into a unified waveform representation, extracts the physiological markers introduced in \autoref{sec:taxonomy}, and concatenates them into a 36-dimensional feature vector. Detailed marker definitions, extraction thresholds, and implementation details are provided in the project repository\footnote{\url{https://github.com/Security-FIT/Physiological-Markers}}.
All recordings are converted to 16\,kHz mono audio and peak-normalized to the $[-1,1]$ range. We then compute a mean energy (RMS) silence mask, which is used for timing-related descriptors such as pauses and active-speech duration.
Respiratory extraction starts from breath intervals detected by \texttt{Respiro-en}~\cite{respiro}. The detected intervals are converted into utterance-level statistics describing breath occurrence, duration, acoustic properties, and speech-breath timing. Vocal extraction estimates $F_0$, perturbation measures, and HNR statistics, using standard acoustic voice-analysis procedures. Temporal extraction uses the silence mask, the Hilbert energy envelope, and Whisper-based filler detection~\cite{whisper} to obtain speech-rate, pause, and filled-pause descriptors. Additional extraction and marker details are available in the project's repository. 
If the extraction of an individual marker fails for a recording, the corresponding value is set to \texttt{NaN}. These missing values are handled in the subsequent classifier processing stage.

% \subsection{Classification Architecture}
% \label{sec:classifier}

\noindent \textbf{Classification Architecture} \quad
Each recording is represented by a 36-dimensional marker vector $\mathbf{x}$. Missing values are imputed using per-marker medians estimated on the training split, and the features are standardized using training-set statistics. The fitted preprocessing is reused unchanged for validation, test, and out-of-domain evaluation.
We train an $\ell_2$-regularized logistic regression (LR) model and an SVM with a radial-basis-function (RBF) kernel, both with balanced class weights. For LR, the deepfake score is $P(y=1\mid\mathbf{x})=\sigma(\mathbf{w}^{\top}\mathbf{x}+b)$, where $y=1$ denotes deepfake speech. Because the markers are standardized, each coefficient $w_j$ describes the change in log-odds associated with a one-standard-deviation increase in marker $j$, holding the remaining markers fixed. For the RBF SVM, the signed decision value is mapped through a sigmoid to obtain a bounded detection score.

\section{Experimental Setup}
\label{sec:setup}

This section describes the data, evaluation protocol, compared systems, and implementation details used to assess the proposed detector based on physiological markers, including cross-corpus generalization, as the proposed markers are intended to capture speech-production cues that are less dependent on a specific synthesis pipeline.

% \subsection{EDS-14 Pooled Dataset}
% \label{sec:datasets}

\noindent \textbf{EDS-14 Pooled Dataset} \quad
To evaluate the proposed detector, we assembled a pooled corpus of English bona fide and deepfake speech drawn from multiple public datasets.
We combine several sources to increase diversity in speakers, recording conditions, and synthesis methods, while restricting the analysis to English recordings with a total duration of at least 14\,s.
% This filtering is necessary because the proposed physiological markers, especially the respiratory and temporal ones, require sufficient recording length for reliable estimation.
We refer to the resulting English Deepfake Speech pooled dataset as EDS-14.
% EDS-14 is built from ASVspoof~5~\cite{wang24_asvspoof}, Codecfake~\cite{codecfake}, In-the-Wild~\cite{muller2022inthewild}, the M-AILABS/MLAAD pair~\cite{MAILABS_2017, muller24_mlaad}, SCDF~\cite{scdf}, and SpeechFake~\cite{speechfake}.
The EDS-14 composition after duration and language filtering is presented in \autoref{tab:eds14}.
The full list of selected recordings included in the EDS-14 dataset is available in the project's repository.
It is important to note that, by pooling multiple datasets, it is not guaranteed that the splits, including the out-of-domain protocol described below, are tool- or speaker-disjoint; i.e., speakers and synthesizers might overlap to some extent across training, validation, and test splits.
This is impossible to guarantee because of varying or missing metadata and labels, as recently described by Staněk et al.~\cite{stanek2026datasets}.

\begin{table}[t]
\centering
\caption{Composition of the aggregated EDS-14 dataset (English recordings longer than 14\,s).}
\label{tab:eds14}
\small
\begin{tabular}{@{}lrr@{}}
\toprule
\textbf{Dataset} & \textbf{Bona fide} & \textbf{Deepfake} \\
\midrule
ASVspoof~5~\cite{wang24_asvspoof} & 9\,208 & 20\,121 \\
Codecfake~\cite{codecfake} & 0 & 79 \\
In-the-Wild~\cite{muller2022inthewild} & 194 & 421 \\
M-AILABS~\cite{MAILABS_2017} + MLAAD~\cite{muller24_mlaad} & 849 & 1\,159 \\
SCDF~\cite{scdf} & 0 & 1\,373 \\
SpeechFake~\cite{speechfake} & 1\,947 & 46\,932 \\
\midrule
\textbf{Total (EDS-14)} & \textbf{12\,198} & \textbf{70\,085} \\
\bottomrule
\end{tabular}
\vspace{-1.3em}
\end{table}

% \subsection{Evaluation Protocol}
% \label{sec:protocol}
\noindent \textbf{Evaluation Protocol} \quad
We evaluate the detector under two complementary settings.
In the \emph{in-domain} (ID) setting, all EDS-14 sources are merged and split at the recording level into training, validation, and test sets with a $70/15/15$ ratio, stratified by label and using a fixed random seed.
The resulting partitions contain distinct recording files, but the same speakers, source utterances, or generators may appear in more than one partition because this information is unavailable for some datasets.
This setting therefore tests performance when training and test recordings come from the same datasets.
To test performance on a different dataset, we additionally use an \emph{out-of-domain} (OOD) setting.
Here, the MLAAD/M-AILABS dataset pair is excluded from training and used only for evaluation.
However, the same speakers, source utterances, or generators may still occur in both groups because this information is unavailable for some datasets~\cite{stanek2026datasets}.
Some component datasets also contain only one class.
We therefore treat the OOD result as performance on a dataset excluded from training, not as performance on guaranteed unseen speakers or generators.
We report equal error rate (EER) as the primary performance metric, computed directly from continuous deepfake scores.

\section{Results}
\label{sec:results}

We report EER, with lower values indicating better performance.
We also inspect the LR coefficients to determine which markers push the score toward bona fide or deepfake speech.

% \subsection{Standalone Physiological-Marker Detection}
% \label{sec:results-standalone}

Table~\ref{tab:standalone} reports the performance of the standalone physiological-marker detector. On the full EDS-14 ID test set, the RBF SVM reaches an EER of $21.40\%$, while LR reaches $27.60\%$. This indicates that the 36 markers capture differences between bona fide and deepfake recordings in the ID test set. With MLAAD/M-AILABS excluded, LR reaches $29.47\%$ EER and the RBF SVM reaches $41.25\%$ on the corresponding reduced ID test set. Because both the training and evaluation composition differ from the full ID setting, this comparison does not isolate the contribution of MLAAD/M-AILABS.
When evaluated on MLAAD/M-AILABS, which was excluded from training, LR reaches $32.77\%$ EER, whereas the RBF SVM reaches $37.83\%$.
Relative to the corresponding reduced ID evaluation, LR deteriorates by $3.30$ percentage points, while the RBF SVM improves by $3.42$ percentage points. The two classifiers therefore behave differently across the evaluations. Overall, LR is more stable across the evaluated settings, while the RBF SVM achieves the best result only on the full ID corpus. These results show that the markers can support detection, but do not show whether the differences arise from physiological properties or from the datasets themselves.

% The OOD setting is more challenging.
% Here, the models trained without MLAAD/M-AILABS are evaluated directly on MLAAD/M-AILABS.
% LR reaches $32.77\%$ EER on the OOD set, whereas the RBF SVM reaches $37.83\%$.
% This degradation shows that a standalone handcrafted-marker detector does not yet generalize reliably to an unseen dataset source.
% At the same time, the OOD EER remains clearly below chance, indicating that physiological markers still provide a measurable detection signal.
% Overall, LR is more stable across the evaluations, while the RBF SVM achieves the best result only in the full ID setting.

\begin{table}[h]
\centering
\caption{Standalone physiological-marker detector EER [\%]. EDS-14$_{\mathrm{all}}$ denotes full EDS-14 and EDS-14$_{\setminus\mathrm{MLAAD}}$ denotes EDS-14 without MLAAD/M-AILABS.}
\label{tab:standalone}
\small
\renewcommand{\arraystretch}{1.15}
% IEEE-only layout: one IEEE column is narrower than the LNI text width, so the
% Train and Eval. columns get a fixed width and wrap their dataset names, and
% the "RBF SVM" header is split into two lines. The table content is unchanged.
\begin{tabular}{@{}l>{\raggedright\arraybackslash}p{0.9in}>{\raggedright\arraybackslash}p{0.9in}cc@{}}
\toprule
\textbf{Setting} & \textbf{Train} & \textbf{Eval.} & \textbf{LR} & \shortstack{\textbf{RBF}\\\textbf{SVM}} \\
\midrule
\multirow{2}{*}{ID}
    & EDS-14$_{\mathrm{all}}$ train & EDS-14$_{\mathrm{all}}$ test & 27.60 & \textbf{21.40} \\
    & EDS-14$_{\setminus\mathrm{MLAAD}}$ train & EDS-14$_{\setminus\mathrm{MLAAD}}$ test & \textbf{29.47} & 41.25 \\
\midrule
OOD
    & EDS-14$_{\setminus\mathrm{MLAAD}}$ train & MLAAD/\newline M-AILABS & \textbf{32.77} & 37.83 \\
\bottomrule
\end{tabular}
\end{table}

\subsection{Intrinsic Marker-Level Explainability}
\label{sec:results-explain}

A key motivation for using physiological markers is that the detector operates in a semantically meaningful feature space.
Unlike conventional SSL-based systems, which output scores from high-dimensional learned representations, the LR model expresses its score as a weighted combination of physiological markers.
The learned coefficients indicate whether and how strongly each standardized marker contributes to the score.
This allows the contribution of each marker to be inspected directly.
The coefficients show how the model separates bona fide and deepfake recordings in the evaluated datasets, but they do not establish causal physiological differences between the classes.

Table~\ref{tab:importance} lists the strongest coefficients of the ID LR model.
Since features are standardized, a positive coefficient indicates that a higher-than-average marker value pushes the score toward the deepfake class, while a negative coefficient indicates that a higher-than-average marker value pushes the score toward the bona fide class.
Lower-than-average values have the opposite effect. The strongest coefficients span all three categories.
Their values may be affected by correlations among markers and differences between datasets and should therefore not be interpreted as causal physiological effects.

\section{Discussion and Conclusion}
\label{sec:conclusion}

We evaluated 36 respiratory, vocal, and temporal markers as a standalone deepfake speech detector.
The best full-ID system reaches $21.40\%$ EER with an RBF SVM, while LR shows more stable performance across the evaluated settings.
These results indicate that physiological markers contain information useful for distinguishing bona fide from deepfake speech in the evaluated datasets.
A key advantage of the LR detector is its interpretability: it expresses each score as a weighted combination of respiratory, vocal, and temporal markers.
The coefficient analysis shows which marker values push the score toward bona fide or deepfake speech.
It does not establish causal physiological differences or show that the same relationships will hold in other datasets.

The approach has several limitations. Recordings must last at least $14\,\mathrm{s}$, which restricts applicability to short-duration scenarios. The same speakers, source utterances, or generators may appear in more than one partition because this information is unavailable for some datasets.
Some component datasets contain recordings from only one class, so the model may learn differences between datasets.
The results are therefore an initial test of the markers, not evidence that they generalize to other datasets. We have also not yet evaluated individual marker groups or their complementarity with modern SSL detectors. Finally, jitter and shimmer remain sensitive to recording quality and extraction errors.

Overall, the proposed approach is an initial step toward biologically grounded deepfake speech detection rather than a replacement for neural detectors.
Future work should examine how speaker identity and utterance duration affect marker distributions and detection performance, using speaker-disjoint and duration-stratified protocols.
It should also improve cross-corpus robustness and marker extraction, for example, through learned markers or explicit breath modeling.
Finally, combining the markers with state-of-the-art SSL-based detectors may improve performance while preserving marker-level interpretability.

% Requires: \usepackage[table]{xcolor}
\newcommand{\dfrow}[1]{\rowcolor{red!#1}}
\newcommand{\bfrow}[1]{\rowcolor{green!#1}}

\begin{table}[t]
\centering
\caption{Most influential markers of the ID LR model, ranked by coefficient magnitude. Red/green indicates that higher-than-average values support the deepfake/bona fide class, respectively.}
\label{tab:importance}
\small
\renewcommand{\arraystretch}{1.15}
\begin{tabular}{llcc}
\toprule
\textbf{Marker} & \textbf{Category} & \textbf{Coeff.} & \textbf{Supports} \\
\midrule
\bfrow{35} Shimmer (APQ5)       & Vocal       & $-4.01$ & Bona fide \\
\dfrow{42} Mean breath duration & Respiratory & $+3.73$ & Deepfake \\
\dfrow{36} Shimmer (absolute)   & Vocal       & $+3.29$ & Deepfake \\
\dfrow{28} Articulation rate    & Temporal    & $+2.51$ & Deepfake \\
\dfrow{18} Jitter (PPQ5)        & Vocal       & $+1.40$ & Deepfake \\
\dfrow{13} Shimmer (APQ11)      & Vocal       & $+0.91$ & Deepfake \\
\bfrow{12} Mean pause duration  & Temporal    & $-0.82$ & Bona fide \\
\bfrow{11} Jitter (RAP)         & Vocal       & $-0.74$ & Bona fide \\
\dfrow{10} Max breath duration  & Respiratory & $+0.68$ & Deepfake \\
\bfrow{8}  Breath frequency     & Respiratory & $-0.50$ & Bona fide \\
\bottomrule
\end{tabular}
\vspace{-1.3em}
\end{table}

\section*{Acknowledgements}

This work was supported by the Brno University of Technology internal project FIT-S-26-9011. Computational resources were provided by the e-INFRA CZ project (ID:90254), supported by the Ministry of Education, Youth and Sports of the Czech Republic. The authors acknowledge the use of generative tools, specifically ChatGPT and Grammarly, for assistance with grammar, paraphrasing, and stylistic refinements.

% IEEE conference papers use BibTeX with the IEEEtran bibliography style.
% The database is shared with the LNI version; only citation formatting differs.
\bibliographystyle{IEEEtran}
\bibliography{lni-paper-example-de}

@article{Firc3,
  title   = {Deepfakes as a threat to a speaker and facial recognition: An overview of tools and attack vectors},
  journal = {Heliyon},
  volume  = {9},
  number  = {4},
  edition = {1},
  pages   = {e15090},
  year    = {2023},
  issn    = {2405-8440},
  doi     = {10.1016/j.heliyon.2023.e15090},
  author  = {Anton Firc and Kamil Malinka and Petr Hanáček}
}

@inproceedings{Babu2021,
  title     = {XLS-R: Self-supervised Cross-lingual Speech Representation Learning at Scale},
  author    = {Arun Babu and Changhan Wang and Andros Tjandra and Kushal Lakhotia and Qiantong Xu and Naman Goyal and Kritika Singh and Patrick {von Platen} and Yatharth Saraf and Juan Pino and Alexei Baevski and Alexis Conneau and Michael Auli},
  year      = {2022},
  booktitle = {Interspeech 2022},
  pages     = {2278--2282},
  doi       = {10.21437/Interspeech.2022-143},
  issn      = {2958-1796},
}

@inproceedings{wang24_asvspoof,
  title     = {{ASVspoof 5: crowdsourced speech data, deepfakes, and adversarial attacks at scale}},
  author    = {Xin Wang and Héctor Delgado and Hemlata Tak and Jee-weon Jung and Hye-jin Shim and Massimiliano Todisco and Ivan Kukanov and Xuechen Liu and Md Sahidullah and Tomi H. Kinnunen and Nicholas Evans and Kong Aik Lee and Junichi Yamagishi},
  year      = {2024},
  booktitle = {{The Automatic Speaker Verification Spoofing Countermeasures Workshop (ASVspoof 2024)}},
  pages     = {1--8},
  doi       = {10.21437/ASVspoof.2024-1},
}

@article{Liu2021,
  author         = {Liu, Tianyun and Yan, Diqun and Wang, Rangding and Yan, Nan and Chen, Gang},
  title          = {Identification of Fake Stereo Audio Using SVM and CNN},
  journal        = {Information},
  volume         = {12},
  year           = {2021},
  number         = {7},
  article-number = {263},
  url            = {https://www.mdpi.com/2078-2489/12/7/263},
  issn           = {2078-2489},
  doi            = {10.3390/info12070263},
  edition        = {1}
}

@inproceedings{muller2022inthewild,
      title={Does Audio Deepfake Detection Generalize?}, 
      author={Nicolas M. Müller and Pavel Czempin and Franziska Dieckmann and Adam Froghyar and Konstantin Böttinger},
      booktitle = {Interspeech},
      year={2022},
}

@inproceedings{tak2022,
  title     = {Automatic Speaker Verification Spoofing and Deepfake Detection Using Wav2vec 2.0 and Data Augmentation},
  author    = {Hemlata Tak and Massimiliano Todisco and Xin Wang and Jee-weon Jung and Junichi Yamagishi and Nicholas Evans},
  year      = {2022},
  booktitle = {The Speaker and Language Recognition Workshop (Odyssey 2022)},
  pages     = {112--119},
  doi       = {10.21437/Odyssey.2022-16},
}

@inproceedings{todisco2019asvspoof,
  title     = {ASVspoof 2019: Future Horizons in Spoofed and Fake Audio Detection},
  author    = {Massimiliano Todisco and Xin Wang and Ville Vestman and Md. Sahidullah and Héctor Delgado and Andreas Nautsch and Junichi Yamagishi and Nicholas Evans and Tomi H. Kinnunen and Kong Aik Lee},
  year      = {2019},
  booktitle = {Interspeech 2019},
  pages     = {1008--1012},
  doi       = {10.21437/Interspeech.2019-2249},
  issn      = {2958-1796},
}

@INPROCEEDINGS{Jung2022aasist,
  author={Jung, Jee-weon and Heo, Hee-Soo and Tak, Hemlata and Shim, Hye-jin and Chung, Joon Son and Lee, Bong-Jin and Yu, Ha-Jin and Evans, Nicholas},
  booktitle={ICASSP 2022 - 2022 IEEE International Conference on Acoustics, Speech and Signal Processing (ICASSP)}, 
  title={AASIST: Audio Anti-Spoofing Using Integrated Spectro-Temporal Graph Attention Networks}, 
  year={2022},
  volume={},
  number={},
  pages={6367-6371},
  doi={10.1109/ICASSP43922.2022.9747766},
}

@inproceedings{rohdin24_asvspoof,
  title     = {BUT systems and analyses for the ASVspoof 5 Challenge},
  author    = {Johan Rohdin and Lin Zhang and Plchot Oldřich and Vojtěch Staněk and David Mihola and Junyi Peng and Themos Stafylakis and Dmitriy Beveraki and Anna Silnova and Jan Brukner and Lukáš Burget},
  year      = {2024},
  booktitle = {The Automatic Speaker Verification Spoofing Countermeasures Workshop (ASVspoof 2024)},
  pages     = {24--31},
  doi       = {10.21437/ASVspoof.2024-4},
}

@INPROCEEDINGS{muller24_mlaad,
  author={Müller, Nicolas M. and Kawa, Piotr and Choong, Wei Herng and Casanova, Edresson and Gölge, Eren and Müller, Thorsten and Syga, Piotr and Sperl, Philip and Böttinger, Konstantin},
  booktitle={2024 International Joint Conference on Neural Networks (IJCNN)}, 
  title={MLAAD: The Multi-Language Audio Anti-Spoofing Dataset}, 
  year={2024},
  volume={},
  number={},
  pages={1-7},
  doi={10.1109/IJCNN60899.2024.10650962}}

@incollection{scdf,
author = "Staněk, Vojtěch and Srna, Karel and Firc, Anton and Malinka, Kamil",
title = "SCDF: A Speaker Characteristics Deepfake Speech Dataset for Bias Analysis",
year = 2025,
doi = "10.18420/biosig_2025_005",
booktitle = "BIOSIG 2025",
publisher = "Gesellschaft für Informatik e.V.",
issn = "2944-7682",
}

@inproceedings{Firc2,
author = {Firc, Anton and Malinka, Kamil},
title = {The dawn of a text-dependent society: deepfakes as a threat to speech verification systems},
year = {2022},
isbn = {9781450387132},
publisher = {Association for Computing Machinery},
address = {New York, NY, USA},
doi = {10.1145/3477314.3507013},
booktitle = {Proceedings of the 37th ACM/SIGAPP Symposium on Applied Computing},
pages = {1646–1655},
numpages = {10},
series = {SAC '22}
}

@InProceedings{MalinkaVoiceAssistants,
author="Kamil {Malinka} and Anton {Firc} and Petr {Kaška} and Tomáš {Lapšanský} and Oskar {Šandor} and Ivan {Homoliak}",
title="Resilience of Voice Assistants to Synthetic Speech",
booktitle="Computer Security -- ESORICS 2024",
year="2024",
publisher="Springer Nature Switzerland",
address="Cham",
pages="66--84",
isbn="978-3-031-70879-4"
}

@inproceedings{elgiriyewithana2024breathing,
  author    = {Nidula Elgiriyewithana and N. D. Kodikara},
  title     = {Attention-Based Efficient Breath Sound Removal in Studio Audio Recordings},
  booktitle = {Proceedings of the 5th International Conference on NLP \& Information Retrieval},
  year      = {2024},
  pages     = {49--58},
  doi       = {10.5121/csit.2024.140604}
}

@inproceedings{xtts,
  title     = {{XTTS: a Massively Multilingual Zero-Shot Text-to-Speech Model}},
  author    = {Edresson Casanova and Kelly Davis and Eren Gölge and Görkem Göknar and Iulian Gulea and Logan Hart and Aya Aljafari and Joshua Meyer and Reuben Morais and Samuel Olayemi and Julian Weber},
  year      = {2024},
  booktitle = {{Interspeech 2024}},
  pages     = {4978--4982},
  doi       = {10.21437/Interspeech.2024-2016},
  issn      = {2958-1796},
}

@INPROCEEDINGS{szekely2020breathing,
  author={Székely, Eva and Henter, Gustav Eje and Beskow, Jonas and Gustafson, Joakim},
  booktitle={ICASSP 2020 - 2020 IEEE International Conference on Acoustics, Speech and Signal Processing (ICASSP)}, 
  title={Breathing and Speech Planning in Spontaneous Speech Synthesis}, 
  year={2020},
  volume={},
  number={},
  pages={7649-7653},
  doi={10.1109/ICASSP40776.2020.9054107}}

@inproceedings{mostaani22_interspeech,
  title     = {{On Breathing Pattern Information in Synthetic Speech}},
  author    = {Zohreh Mostaani and Mathew Magimai Doss},
  year      = {2022},
  booktitle = {{Interspeech 2022}},
  pages     = {2768--2772},
  doi       = {10.21437/Interspeech.2022-10271},
  issn      = {2958-1796},
}

@article{fantini2024voice,
author = {Fantini, Marco and Ciravegna, Gabriele and Koudounas, Alkis and Cerquitelli, Tania and Baralis, Elena and Succo, Giovanni and Crosetti, Erika},
year = {2024},
month = {11},
pages = {},
title = {The Rapidly Evolving Scenario of Acoustic Voice Analysis in Otolaryngology},
volume = {16},
journal = {Cureus},
doi = {10.7759/cureus.73491}
}

@article{deJong2021fluency,
author = {Nivja H. de Jong and Jos Pacilly and Willemijn Heeren},
title = {PRAAT scripts to measure speed fluency and breakdown fluency in speech automatically},
journal = {Assessment in Education: Principles, Policy \& Practice},
volume = {28},
number = {4},
pages = {456--476},
year = {2021},
publisher = {Routledge},
doi = {10.1080/0969594X.2021.1951162},
URL = {https://doi.org/10.1080/0969594X.2021.1951162},
}

@Article{angelopoulou2024fluency,
AUTHOR = {Angelopoulou, Georgia and Kasselimis, Dimitrios and Goutsos, Dionysios and Potagas, Constantin},
TITLE = {{A Methodological Approach to Quantifying Silent Pauses, Speech Rate, and Articulation Rate across Distinct Narrative Tasks: Introducing the Connected Speech Analysis Protocol (CSAP)}},
JOURNAL = {Brain Sciences},
VOLUME = {14},
YEAR = {2024},
NUMBER = {5},
ARTICLE-NUMBER = {466},
PubMedID = {38790445},
ISSN = {2076-3425},
DOI = {10.3390/brainsci14050466}
}

@INPROCEEDINGS{tran2023breath,
  author={Tran, Van-Thuan and You, Ting-Hao and Tsai, Wei-Ho},
  booktitle={2023 International Conference on Artificial Intelligence and Power Engineering (AIPE)}, 
  title={Breath Sound-based Person Identification in Noisy Conditions}, 
  year={2023},
  volume={},
  number={},
  pages={50-55},
  doi={10.1109/AIPE58786.2023.00016}}

@article{Li2020Breathing,
author = {Lu, Li and Liu, Lingshuang and Hussain, Muhammad Jawad and Liu, Yongshuai},
title = {I Sense You by Breath: Speaker Recognition via Breath Biometrics},
year = {2020},
issue_date = {March-April 2020},
publisher = {IEEE Computer Society Press},
address = {Washington, DC, USA},
volume = {17},
number = {2},
issn = {1545-5971},
url = {https://doi.org/10.1109/TDSC.2017.2767587},
doi = {10.1109/TDSC.2017.2767587},
journal = {IEEE Transactions on Dependable and Secure Computing},
month = mar,
pages = {306–319},
numpages = {14}
}

@misc{stanek2026evolutionary,
      title={Evolutionary Multi-Objective Fusion of Deepfake Speech Detectors}, 
      author={Vojtěch Staněk and Martin Perešíni and Lukáš Sekanina and Anton Firc and Kamil Malinka},
      year={2026},
      eprint={2604.01330},
      archivePrefix={arXiv},
      primaryClass={cs.SD},
      url={https://arxiv.org/abs/2604.01330},
}

@inproceedings{respiro,
  title     = {{Frame-Wise Breath Detection with Self-Training: An Exploration of Enhancing Breath Naturalness in Text-to-Speech}},
  author    = {Dong Yang and Tomoki Koriyama and Yuki Saito},
  year      = {2024},
  booktitle = {{Interspeech 2024}},
  pages     = {4928--4932},
  doi       = {10.21437/Interspeech.2024-168},
  issn      = {2958-1796},
}

@inproceedings{whisper,
author = {Radford, Alec and Kim, Jong Wook and Xu, Tao and Brockman, Greg and McLeavey, Christine and Sutskever, Ilya},
title = {Robust speech recognition via large-scale weak supervision},
year = {2023},
publisher = {JMLR.org},
booktitle = {Proceedings of the 40th International Conference on Machine Learning},
articleno = {1182},
numpages = {27},
location = {Honolulu, Hawaii, USA},
series = {ICML'23}
}

@misc{MAILABS_2017,
   author = {Solak, I. Celeste Aurora and Naumov, Dima},
   title = {The M-AILABS Speech Dataset},
   year = {2017},
   howpublished = {\url{https://github.com/i-celeste-aurora/m-ailabs-dataset}}
}

@inproceedings{speechfake,
    title = "{S}peech{F}ake: A Large-Scale Multilingual Speech Deepfake Dataset Incorporating Cutting-Edge Generation Methods",
    author = "Huang, Wen  and
      Gu, Yanmei  and
      Wang, Zhiming  and
      Zhu, Huijia  and
      Qian, Yanmin",
    editor = "Che, Wanxiang  and
      Nabende, Joyce  and
      Shutova, Ekaterina  and
      Pilehvar, Mohammad Taher",
    booktitle = "Proceedings of the 63rd Annual Meeting of the Association for Computational Linguistics (Volume 1: Long Papers)",
    month = jul,
    year = "2025",
    address = "Vienna, Austria",
    publisher = "Association for Computational Linguistics",
    url = "https://aclanthology.org/2025.acl-long.493/",
    doi = "10.18653/v1/2025.acl-long.493",
    pages = "9985--9998",
    ISBN = "979-8-89176-251-0",
}

@inproceedings{codecfake,
  title     = {{Codecfake: An Initial Dataset for Detecting LLM-based Deepfake Audio}},
  author    = {Yi Lu and Yuankun Xie and Ruibo Fu and Zhengqi Wen and Jianhua Tao and Zhiyong Wang and Xin Qi and Xuefei Liu and Yongwei Li and Yukun Liu and Xiaopeng Wang and Shuchen Shi},
  year      = {2024},
  booktitle = {{Interspeech 2024}},
  pages     = {1390--1394},
  doi       = {10.21437/Interspeech.2024-2428},
  issn      = {2958-1796},
}

@inproceedings{stanek2026datasets,
  title     = {{Ethical and Technical Limits of Deepfake Speech Datasets}},
  author    = {Vojtěch Staněk and Eva Trnovská and Kamil Malinka and Anton Firc},
  year      = {2026},
  booktitle = {{Interspeech 2026}},
  pages     = {2340--2345},
  doi       = {10.21437/Interspeech.2026-124},
  issn      = {2958-1796},
}

@article{layton_breath,
author = {Layton, Seth and others},
title = {Every Breath You Don’t Take: Deepfake Speech Detection Using Breath},
year = {2025},
issue_date = {September 2025},
publisher = {Association for Computing Machinery},
address = {New York, NY, USA},
volume = {6},
number = {3},
url = {https://doi.org/10.1145/3754456},
doi = {10.1145/3754456},
journal = {Digital Threats},
month = sep,
articleno = {13},
numpages = {18}
}

@misc{breathnet,
      title={BreathNet: Generalizable Audio Deepfake Detection via Breath-Cue-Guided Feature Refinement}, 
      author={Zhe Ye and Xiangui Kang and Jiayi He and Chengxin Chen and Wei Zhu and Kai Wu and Yin Yang and Jiwu Huang},
      year={2026},
      eprint={2602.13596},
      archivePrefix={arXiv},
      primaryClass={cs.SD},
      url={https://arxiv.org/abs/2602.13596}, 
}

\end{document}